\documentclass[a4paper]{article}
\usepackage{subfig}

\usepackage{INTERSPEECH2020}

\title{XiaoiceSing: A High-Quality and Integrated Singing Voice Synthesis System}
\name{Peiling Lu$^1$, Jie Wu$^1$, Jian Luan$^1$, Xu Tan$^2$, Li Zhou$^1$}
\address{
  $^1$Xiaoice, Microsoft Software Technology Center Asia\\
  $^2$Microsoft Research Asia}
\email{\{peil, jiewu, jianluan, xuta, lzhou\}@microsoft.com}
\begin{document}

\maketitle
\begin{abstract}
  This paper presents XiaoiceSing, a high-quality singing voice synthesis system which employs an integrated network for spectrum, F0 and duration modeling. We follow the main architecture of FastSpeech while proposing some singing-specific design: 1) Besides phoneme ID and position encoding, features from musical score (e.g.note pitch and length) are also added. 2) To attenuate off-key issues, we add a residual connection in F0 prediction. 3) In addition to the duration loss of each phoneme, the duration of all the phonemes in a musical note is accumulated to calculate the syllable duration loss for rhythm enhancement. Experiment results show that XiaoiceSing outperforms the baseline system of convolutional neural networks by 1.44 MOS on sound quality, 1.18 on pronunciation accuracy and 1.38 on naturalness respectively. In two A/B tests, the proposed F0 and duration modeling methods achieve 97.3\% and 84.3\% preference rate over baseline respectively, which demonstrates the overwhelming advantages of XiaoiceSing.
\end{abstract}

\section{Introduction}

Singing voice synthesis (SVS) is an attracting technique to generate singing voices from musical score information (e.g., lyrics, tempo and pitch). In order to generate explicit, expressive and rhythmic singing voice of good quality, three aspects should be taken into consider: 1) A powerful spectrum model that predicts spectral features for articulate pronunciation, adequate sound quality and well-behaved naturalness. 2) An effective F0 model to establish the complex patterns in F0 contour of singing. 3) A duration model that can learn the rhythmic factors of singing.

Traditional methods in SVS including unit concatenation \cite{gu2016singing, bonada2016expressive, kenmochi2007vocaloid, macon1997concatenation} and statistical parametric synthesis \cite{saino2006hmm, nakamura2014hmm} have been dominant for many years. However, the performance of sound quality, pronunciation, and naturalness still has a large gap to reach that of natural recording. Recently, many deep learning based models such as feed-forward neural neworks (FFNN) \cite{nishimura2016singing} and long short-term memory (LSTM) \cite{kim2018korean} are introduced into SVS for improving the acoustic model. Generative adversarial network (GAN) \cite{hono2019singing} is also applied to feed-forward neural network for alleviating the over-smoothing effect. And a senquence-to-sequence system using a feed-forward transformer (FFT) \cite{blaauw2020sequence} shows the ability of avoiding the exposure bias issues when modeling timbre, while another convolutional neural network described by \cite{nakamura2019fast} can model the long-term dependencies of singing voices. More advanced networks like WaveNet is introduced to predict acoustic features \cite{blaauw2017neural} and is able to capture the characteristic of singing voices more precisely. Moreover, deep autoregressive neural network is proposed in \cite{yi2019singing} and \cite{lee2019adversarially} to model spectral features. However, lots of efforts in these work are made on spectral features, ignoring the dynamic and rhythmic impact brought about by F0 and duration.

Some research also attempt to ensure the precision of F0 and duration in SVS. For F0 modeling, a strategy that computing a weighted average of predicted F0 over note is proposed in \cite{blaauw2017neural}. Performing a moving average on predicted F0 contour is also considered effective in \cite{yi2019singing}. This implies that it is still a challenge for model itself to alleviate the out-of-tune issue of synthesized singing voice. And the post-processing strategies may smooth the dynamics of F0 contour, and neglect the potential interactions between F0 and spectral features. As for duration model, to synchronize the predicted duration with musical note duration, fitting heuristic strategy in \cite{kim2018korean} confirms the total of predicted phoneme duration to note duration, while normalizing the predicted duration within the sum of phoneme duration is used as post-processing method in \cite{blaauw2017neural}. It shows that accurately predicting duration by model itself is still an important issue for SVS. Furthermore, in the systems mentioned above, duration, F0 and spectrum models are trained independently, which usually causes the problem that their consistency in singing voices and the dependencies of musical features are ignored. 

Our proposed system XiaoiceSing is inspired by FastSpeech~\cite{ren2019FastSpeech}, a significant work that achieves high-quality and fast-speed voice generation in Text-to-Speech (TTS). To suit SVS task, we adopt FastSpeech network with some modifications in the following aspects: 1) Besides phoneme sequence of lyrics, all musical score information (e.g., note duration, note pitch) is encoded as input. 2) To avoid out-of-tune issue, we add a residual connection between note pitch and predicted F0, 3) In addition to phoneme duration loss, syllable duration loss is taken into account during training for rhythm enhancement. 4) We model vocoder features including mel-generalized cepstrum (MGC) and band aperiodicity (BAP) rather than mel-spectrogram, and use WORLD vocoder \cite{morise2016world} to synthesize singing voice, which can guarantee the consistency between the input F0 contour and that of generated singing voice. In general, spectrum, F0 and duration model are trained jointly by utilizing the structure in Fastspeech that decoder and duration predictor shares the same encoder.

\begin{figure}[htb]
\vspace{-15pt}
\begin{minipage}[b]{1.0\linewidth}
  \centering
  \centerline{\includegraphics[width=6.5cm]{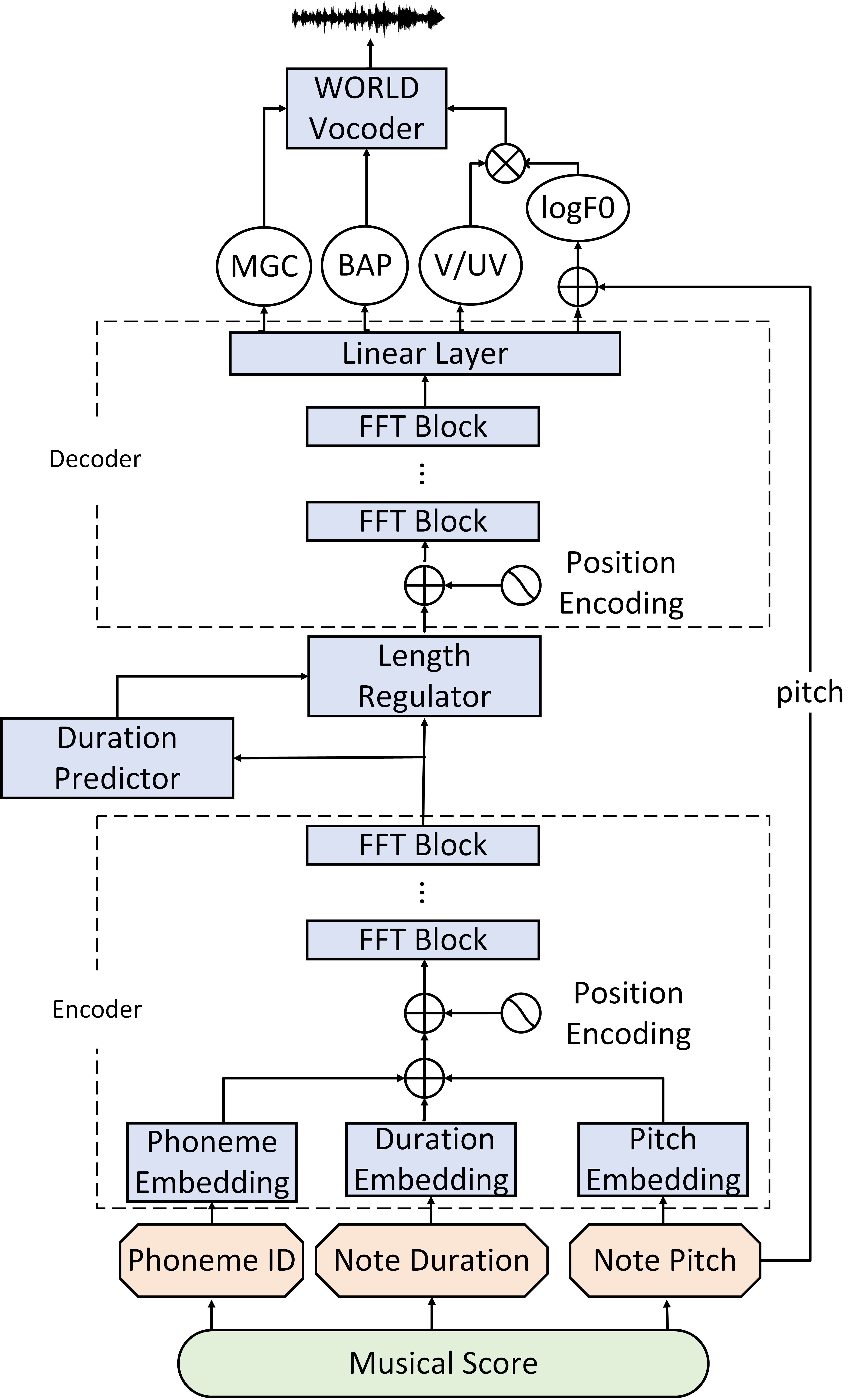}}
\end{minipage}
\caption{Architecture of XiaoiceSing based on modified FastSpeech}
\label{fig:struct}
\vspace{-18pt}
\end{figure}

\section{Architecture}
\label{sec:format}

As illustrated in Figure~\ref{fig:struct}, the proposed XiaoiceSing system integrates F0 and duration model with spectrum model based on a modified FastSpeech network and then generates waveform through WORLD vocoder. This network  consists of several modules: 1) A musical score encoder to convert phoneme name, note duration and pitch sequence into a dense vector sequence. 2) A duration predictor to get phoneme duration from the encoder vector sequence, and a length regulator to expand encoder vector sequence according to the predicted phoneme duration. 3) A decoder to generate acoustic features from the expanded encoded vectors.

\subsection{Musical score encoder}

\begin{figure}[t!]
\vspace{-15pt}
\begin{minipage}[b]{1.0\linewidth}
  \centering
  \centerline{\includegraphics[width=8cm]{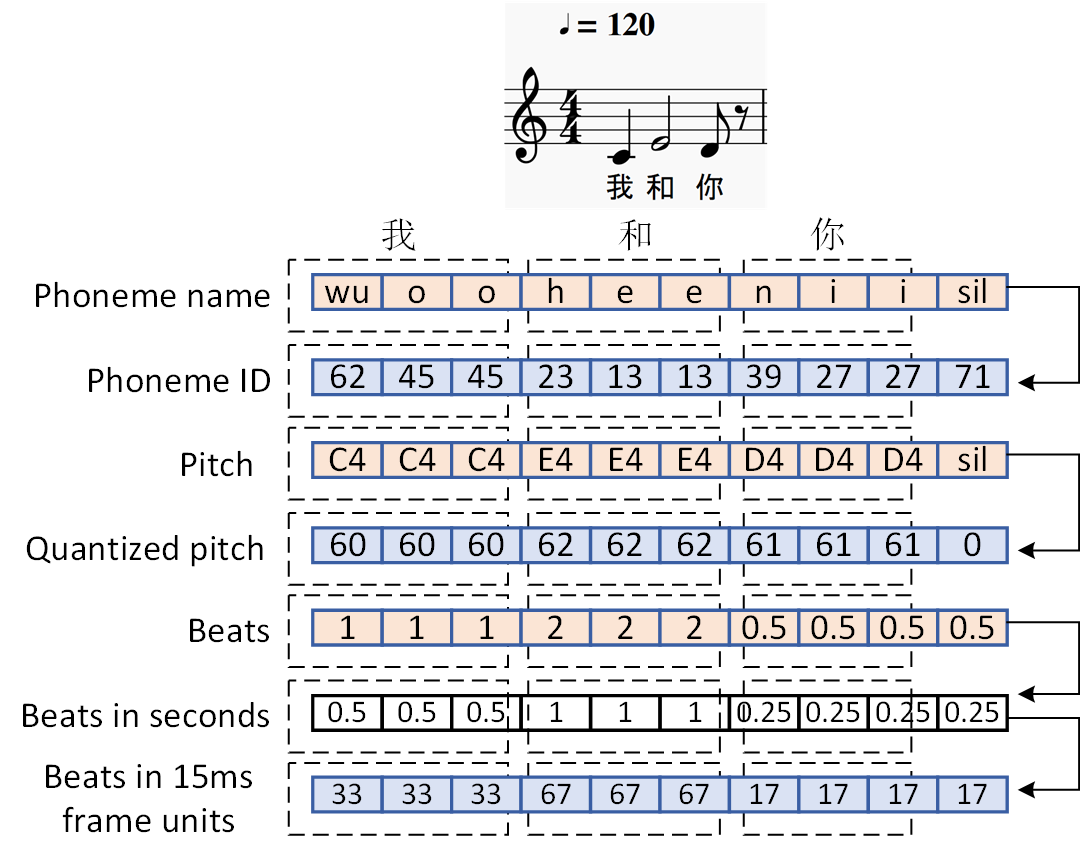}}
\vspace{-5pt}
\end{minipage}
\caption{Musical score representation}
\label{fig:input}
\vspace{-15pt}
\end{figure}

Generally, musical score contains lyrics, note pitch and note duration, which are necessary to sing a song. As shown in Figure~\ref{fig:input}, we first convert lyrics into phoneme sequence with grapheme to phoneme conversion \cite{taylor2005hidden}, where each syllable in the lyrics is decomposed into multiple phonemes. Each note pitch is converted into a pitch ID following the MIDI standard \cite{midi}, while the note duration is quantified by music tempo and converted into frame count (the number of phoneme frames). The note pitch and note duration are duplicated to fit the length of  phonemes. Therefore, the musical score input $S \in \mathbb{R}^{N \times 3}$ is in phoneme level, where $N$ denotes the number of phonemes. For each phoneme, the input feature includes its phoneme ID, the corresponding note pitch and duration. They are embedded separately into a dense vector in the same dimension and then added together with position encoding. The resulting vector is then passed to the encoder,  which contains multiple FFT blocks. Each FFT block consists of a self-attention network and a two-layer 1-D convolution network with ReLU activation~\cite{ren2019FastSpeech}.

\subsection{Duration predictor}
The phoneme duration predictor consists of a 1-D convolutional network and is trained to guide the sequence expansion in length regulator, which follows the structure in~\cite{ren2019FastSpeech}. Note that the duration predictor leverages the same encoder output with spectrum and F0 prediction. Research in \cite{deutsch2013psychology} shows that rhythm is based on an order which is primarily temporal. Besides phoneme duration, syllable duration also plays an important role in the rhythm of SVS. Only focusing on learning phoneme level duration is not enough to achieve good rhythmic pattern. Therefore, we propose to add a control for syllable-level duration as well. Specifically, the note is always correlated to syllable in lyrics. One syllable may corresponds to one or more notes. Thus, a syllable-level duration loss $L_{sd}$ between the ground-truth syllable duration and the sum of predicted duration of all phonemes in the syllable is designed to strengthen their relationship. The loss of duration predictor $L_{dur}$ is defined as below:
\begin{equation} \label{eq2}
\begin{split}
L_{dur} & = w_{pd} * L_{pd} + w_{sd} * L_{sd}
\end{split}
\end{equation}
where $L_{pd}$ and $L_{sd}$ represent the loss of phoneme duration and syllable duration respectively, while $w_{pd}$ and $w_{sd}$ are the corresponding weights.

\subsection{Decoder}
In the proposed system, WORLD vocoder is used to generate waveform, since it has explicit F0 control and can guarantee the correct expression of extremely high and low tones. Thus, the decoder will predict MGC and BAP features instead of mel-spectrogram. The loss for spectral parameters $L_{spec}$ is calculated as follows:
\begin{equation} \label{eq1}
\begin{split}
L_{spec} & = w_{m} * L_{m} + w_{b} * L_{b}
\end{split}
\end{equation}
where, $L_{m}$ and $L_{b}$ mean the loss of MGC and BAP respectively, while $w_{m}$ and $w_{b}$ are the corresponding weights.

Compared to speech, singing has more complicated and sensitive F0 contour. For example, it has a wider range from 80 to 3400 Hz. And the dynamics of F0 movements like vibrato and overshoot helps to convey emotion more expressively. It is also pointed out in \cite{deutsch2013psychology} that a little deviation from the standard pitch will impair the listening experience a lot. On the other hand, the training data can hardly cover all the pitch range with sufficient cases. That means, the F0 prediction may have issues if the input note pitch is not shown or rare in training data. Data augmentation may address this problem by implementing pitch shift on training data. But it's not economy and may cause much longer training time. Instead, we propose a residual connection between input and output pitch (here we use logF0, the log scale of F0), as shown in Figure~\ref{fig:struct}. In this way, the decoder only need to predict the human bias from the standard note pitch, which is more robust to  rare or unseen data. Later experiments also prove its advantages. As usual, F0 prediction is accompanied by V/UV decision. Since V/UV decision is binary, logistic regression is applied here. Eventually, the loss function for decoder output $L_{dec}$ is revised as:
\begin{equation} \label{eq3}
\begin{split}
L_{dec} & = L_{spec} + w_{f} * L_{f} + w_{u} * L_{u}
\end{split}
\end{equation}
where, $L_{f}$ and $L_{u}$ indicates the loss of logF0 and V/UV decision separately, while $w_{f}$ and $w_{u}$ is the corresponding weights.

\section{Experiments}
\label{sec:pagestyle}

\subsection{Dataset}
2297 Mandarin pop songs are collected from a female singer in a professional recording room, where the singer sings along with the accompaniment audio. After that, the musical score is manually reviewed according to the recording. All the audio are sampled at 48kHz with 16-bit quantization and split into segments whose length is shorter than 10 seconds. Finally, 10761 segments, about 74 hours  data, are obtained in total. Among them, 9685 pieces are randomly selected for training, while the rest 1076 for validation and test. From recording, acoustic features are extracted by WORLD with 15ms frame shift, including 60-d MGC, 5-d BAP, 1-d logF0, and 1-d V/UV flag. The phoneme duration label is obtained by HMM-based forced alignment \cite{sjolander2003hmm}.

\subsection{System configuration}
The following two systems are constructed for evaluating the performances of our proposed systems.
\begin{itemize}
\item Baseline: convolutional neural network (CNN) based architecture for spectral parameters prediction, combining with independent LSTM-based F0 and duration model.
\item XiaoiceSing: Modified FastSpeech architecture to predict spectral parameters, F0 and duration at the same time.
\end{itemize}

The baseline system\cite{nakamura2019fast} contains an acoustic model to predict MGC, BAP and V/UV. Meanwhile, independent F0 and duration model are trained separately. In acoustic model, the encoder consists of 3 hidden layers of feed-forward neural network. A length regulator following copies the encoder output into frame lengths. And the decoder is composed of several convolutional layers, with 2 layers for down-sampling, 6 layers of residual structure, and 2 layers for up-sampling. The F0 and duration model has the same network structure of one feed-forward layer and two LSTM layers. In phoneme duration model, 132-d note level features (e.g., note duration, note pitch) are calculated from musical score as input. Its output is 3-d phoneme duration vector of each syllable. In F0 model, input features are at frame level, including 800-d linguistic feature (e.g., the phoneme identity with context), 55-d musical score information (e.g., note duration, note pitch and frame position).


In the proposed system XiaoiceSing, the encoder and decoder both stack 6 FFT blocks. The size of the phoneme vocabulary is 72. The encoder converts input musical score into a 384-d vector, while the decoder outputs 67-d acoustic features. The loss of MGC, BAP and logF0 are computed separately with L1 regularization loss function. Only V/UV decision utilizes binary cross entropy as the loss function. The duration predictor with 2-layer 1-D convolutional network is jointly trained. L1 regulation loss function is applied for both phoneme and note duration loss calculation. The other hyper-parameters in network are set following the same configuration in~\cite{ren2019FastSpeech}. Adam optimizer with $\beta1=0.9, \beta2=0.98, \epsilon=10^{-9}$ is selected for optimization strategy. We train our model on one NVIDIA P100 GPU with the batch size of 32 sentences and follows the learning rate in \cite{vaswani2017attention}. It takes 40k iteration steps to converge.

\subsection{Overall performance}

\begin{table}[t!]
\caption{Mean opinion scores (MOS) test results}
\vspace{-5pt}
\hfill \break
 \begin{tabular}{c c c c} 
 \toprule
   \textbf{Evaluation Item}& \textbf{Recording} & \textbf{XiaoiceSing} & \textbf{Baseline} \\ [0.5ex] 
\midrule
 Pronun acc. &  $4.79\pm0.46$ & $4.32\pm0.78$ & $3.14\pm1.10$ \\ 

Sound quality & $4.54\pm0.60$ & $3.70\pm0.84$ & $2.26\pm0.95$ \\

 Naturalness & $4.78\pm0.45$ & $3.61\pm0.77$ & $2.23\pm0.93$ \\
\bottomrule
\end{tabular}
\label{tab:mos}
\end{table} 

\begin{table}[t!]
\caption{Objective evaluation results of different systems.}
\vspace{-5pt}
\setlength{\tabcolsep}{6mm}
\hfill \break
\begin{tabular}{c c c} 
\toprule
    \textbf{Evaluation Item}& \textbf{XiaoiceSing} & \textbf{Baseline} \\  [0.5ex] 
\midrule
Dur RMSE & $20.55$ & $24.39$\\

Dur CORR & $0.91$ & $0.69$\\ 

F0 RMSE (Hz) &  $10.45$ & $13.74$\\ 

F0 CORR & $0.99$ & $0.91$\\

MCD (dB) & $5.42$ & $6.09$\\

BAPD (dB) & $25.12$ & $28.39$\\

V/UV Error (\%) & $2.66$ & $4.27$\\

\bottomrule
\end{tabular}
\label{tab:obj}
\vspace{-13pt}
\end{table}

In this experiment, baseline and XiaoiceSing systems are compared to evaluate the overall performance.

We first conduct mean opinion score (MOS) test for the two systems. For each system, we prepare 30 audio samples, each sample within about 10 seconds. We ask 10 listeners to rate the MOS score of each sample according to three aspects: pronunciation accuracy, sound quality and naturalness. The MOS score is averaged over all the samples to get the final score.

As shown in Table~\ref{tab:mos}, XiaoiceSing outperforms baseline in all three aspects. Specifically, XiaoiceSing attains a MOS score of 4.32 on pronunciation accuracy, which is close to 4.79 of recording, and 1.18 over baseline. On sound quality and naturalness, XiaoiceSing also achieves a gain of 1.44 and 1.38 over baseline respectively. Although MOS score of XiaoiceSing drops down below 4.0 on these two items, it is likely due to some deficiencies of WORLD or higher expectations of listeners. Additionally, XiaoiceSing has less standard deviations on MOS than baseline, which may indicates that self-attention mechanism brings about the better performance stability among all test cases.

We further calculate some objective metrics including duration root mean square error (Dur RMSE), duration correlation (Dur CORR), F0 RMSE, F0 CORR, mel-cepstral distortion (MCD), band aperiodicity distortion (BAPD) and voiced/unvoiced error rate (V/UV error) to measure the quality of synthesized singing voice. The evaluation results are listed in Table~\ref{tab:obj}. It shows that XiaoiceSing achieves lower RMSE and higher CORR both in duration and F0 than baseline. Meanwhile, MCD, BAPD and V/UV error of XiaoiceSing are all lower than those of baseline. We can see that XiaoiceSing has a stronger ability to generate accurate phoneme duration, F0 and spectral features.

\begin{figure}[t!]
\begin{minipage}[b]{1.0\linewidth}
  \centering
  \centerline{\includegraphics[width=7cm]{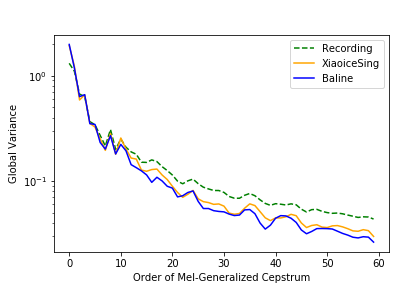}}
\end{minipage}
\caption{Averaged GVs of Mel-Generalized Cepstrum coefficients.}
\label{fig:gv}
\vspace{-10pt}
\end{figure}

\begin{figure}[t!]
    \centering
    \subfloat[]{\includegraphics[width=0.2\textwidth]{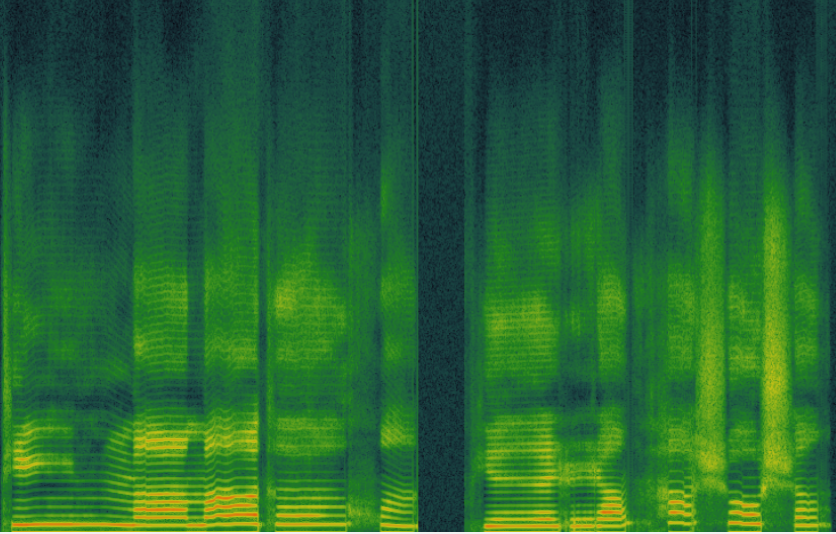}}
    \subfloat[]{\includegraphics[width=0.2\textwidth]{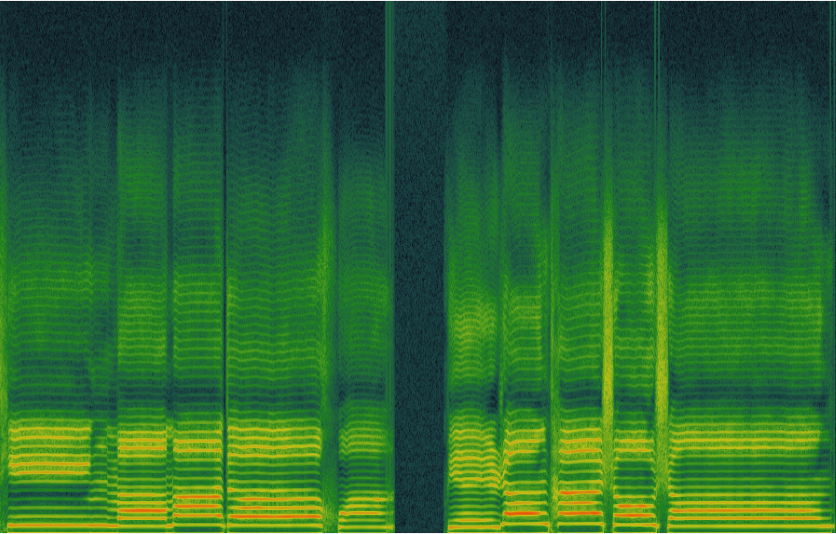}} 
    \caption{(a) mel-spectrum of samples predicted by Baseline. (b) mel-spectrum of samples predicted by XiaoiceSing.}
    \label{fig:mel}
\vspace{-15pt}
\end{figure}

We also calculate the averaged global variance (GV) of mel-generalized cepstrum coefficients for each system in Figure~\ref{fig:gv}. It is shown that XiaoiceSing improves the averaged GV over the baseline especially in high frequency level, proving a better performance in stability. This can also be demonstrated in Figure~\ref{fig:mel}. We plot the mel-spectrum of the audio samples generated by two systems. It is clear that the sample generated by XiaoiceSing has a more solid and distinct shape in high frequency components.

\subsection{Evaluation on F0 modeling}
In this experiment, we compare the F0 model in XiaoiceSing with a seperate LSTM-based F0 model as used in baseline, to validate the effectiveness of proposed F0 modeling.

We conducted an A/B test for evaluating the correctness and dynamics of F0 contour, the listeners are asked to listen to the recording as reference before they compare the same song generated by these two approaches in a random order. The preference is selected by judging which one is more similar with recording on pitch perception. As shown in the first bar of Figure~\ref{fig:abtest}, 97.3\% supports XiaoiceSing, while only 1.70\% supports baseline.

\begin{figure}[t!]
\begin{minipage}[b]{1.0\linewidth}
  \centering
  \centerline{\includegraphics[width=7cm]{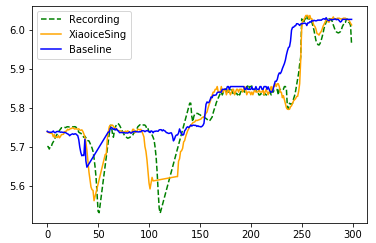}}
\end{minipage}
\vspace{-10pt}
\caption{F0 contour comparison for different systems.}
\label{fig:freq}
\vspace{-12pt}
\end{figure}

To compare the F0 modeling performance objectively, we plot the the same part of F0 contour generated by different approaches in Figure~\ref{fig:freq}. Although both systems can express the pitch of notes correctly, F0 predicted from LSTM model is over-smoothed and cannot express the dynamics of F0 contour like vibratos. XiaoiceSing seems much better and closer to the recording.

\subsection{Evaluation on duration modeling}
In this experiment, we compare the duration model in XiaoiceSing with a separate LSTM-based duration model to evaluate the advantage of the proposed duration modeling.

\begin{figure}[t!]
\begin{minipage}[b]{1.0\linewidth}
  \centering
  \centerline{\includegraphics[width=7.5cm]{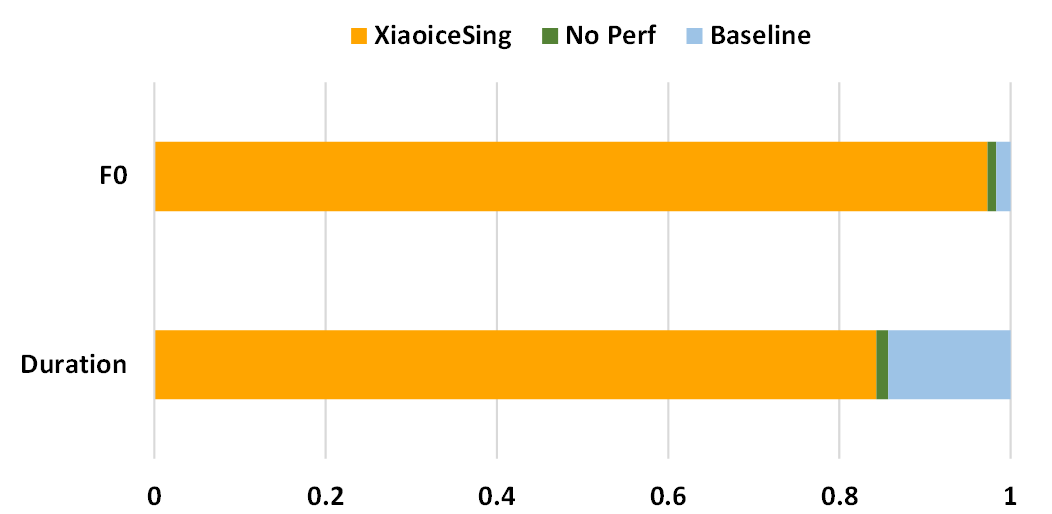}}
\end{minipage}
\vspace{-8pt}
\caption{A/B preference test results for F0 and duration.}
\label{fig:abtest}
\vspace{-5pt}
\end{figure}

Similarly, another A/B test is carried out to evaluate the duration of rhythm preference, as shown in the second bar of Figure~\ref{fig:abtest}. It shows 84.3\% supports XiaoiceSing, while only 14.3\% supports baseline.

\begin{figure}[t!]
\begin{minipage}[b]{1.0\linewidth}
  \centering
  \centerline{\includegraphics[width=8cm]{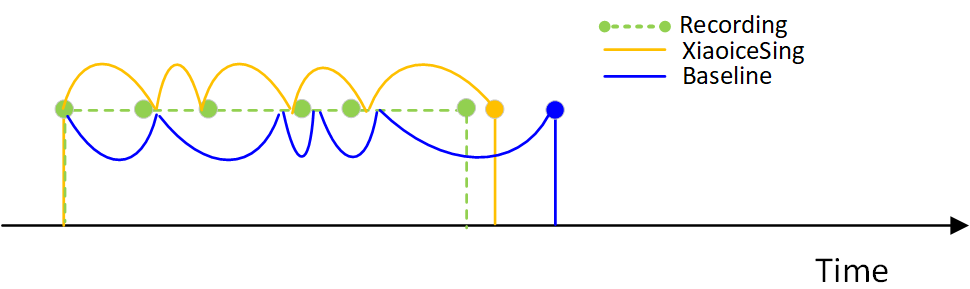}}
\end{minipage}
\vspace{-15pt}
\caption{Phoneme duration comparison for different systems.}
\label{fig:dur}
\vspace{-15pt}
\end{figure}

The duration of five consecutive phonemes are compared in Figure~\ref{fig:dur}. They are left aligned and plotted along time axis. Dots on the green line indicates start time and end time of each phoneme in recording. It is clear that XiaoiceSing predicts the duration more closely to recording. A separate LSTM-based duration model is quite unstable and predicts a very long duration for the last phoneme. Moreover, its accumulated errors is much greater than XiaoiceSing.

Results of the two A/B tests in the last two subsections approve the great advantage of our proposed system on duration and F0 prediction. It is consistent with the conclusion of above objective evaluation. Some samples for the subjective listening test are available via this link\footnote{https://xiaoicesing.github.io/}.

\section{Conclusions}
\label{sec:typestyle}

In this paper, we presented a high-quality singing voice synthesis system XiaoiceSing. With stacked CNN and attention network, the remote dependencies along time axis can be better modeled. And the integrated model ensures the consistency among predicted acoustic features. Experiments results demonstrate that it achieves great advantages on sound quality, pronunciation accuracy, and naturalness over the baseline system. Especially, the performances on F0 perception are outstanding because of the residual connection between note pitch and predicted F0, while the improvements on duration prediction are remarkable due to the added syllable duration constraint between expected syllable duration and predicted phone durations.


\bibliographystyle{IEEEtran}

\bibliography{mybib}


\end{document}